\documentstyle[twocolumn,seceq,epsfig]{jpsj}

\def\runtitle{Least Action Principle for the Real-Time DMRG}
\def\runauthor{Kouji {\sc Ueda}, Chenglong {\sc Jin}, Naokazu {\sc Shibata}, 
Yasuhiro {\sc Hieida}, and Tomotoshi {\sc Nishino}}

\title{Least Action Principle for the Real-Time Density Matrix Renormalization Group}

\author
{Kouji {\sc Ueda}, Chenglong {\sc Jin}, Naokazu {\sc Shibata}$^{2)}_{~}$, 
Yasuhiro {\sc Hieida}$^{3)}_{~}$, and Tomotoshi {\sc Nishino}}

\inst
{Department of Physics, Faculty of Science, Kobe University, Kobe 657-8501\\
$^{(2)}_{~}$ 
Department of Physics, Graduate School of Science, Tohoku University, Sendai 980-8578\\
$^{(3)}$Computer and Network Center, Saga University, Saga 840-8502}

\abst
{A kind of least action principle is introduced for the discrete time evolution of one-dimensional quantum 
lattice models. Based on this principle, we obtain an optimal condition 
for the matrix product states on succeeding time slices generated by the real-time density 
matrix renormalization group method. This optimization can also be applied to classical
simulations of quantum circuits. We discuss the time reversal symmetry in the 
fully optimized MPS.}

\kword{DMRG, Time Evolution, Least Action, MPS}

\begin{document}
\sloppy
\maketitle

\section{Introduction}

The density matrix renormalization group (DMRG) method has been widely applied
to calculations of eigenstates of low-dimensional quantum 
systems.~\cite{White,Peschel,Shoe} The method can be regarded as a numerical
variational method, which optimizes position dependent matrix product state (MPS) by way
of iterative improvements of local matrices.~\cite{Ostlund,Takasaki} This variational
background guarantees that truncation error in the block spin transformation
does not accumulate in the iterative numerical calculations of the finite system 
DMRG algorithm.

One of the recent progress in DMRG is its application to quantum states under real or 
imaginary time evolution.~\cite{Cazalilla,Xiang,Vidal,D_White,D_Shoe} 
The concept of adopted time 
dependence is a key feature in the real-time DMRG method,
where the evolving quantum state is approximated by 
MPS as precise as possible at each time slice.~\cite{Vidal,D_White,D_Shoe,Verst_1}
In this article we focus on the weak breaking of the time-reversal symmetry 
in the numerical algorithm of the real-time DMRG method.
Suppose that we start from an initial state $| \Psi( t_I^{~} ) \rangle$.
After the numerical time evolution with respect to the 
time-independent Hamiltonian $H$, we 
get to the calculated final state $| \Psi( t_F^{~} ) \rangle$ that approximates 
$\exp[ ( t_F^{~} - t_I^{~} ) H / i \hbar ] | \Psi( t_I^{~} ) \rangle$. 
The backward numerical time evolution from $| \Psi( t_F^{~} ) \rangle$ toward
past direction 
gives the calculated state $| \Psi'( t_I^{~} ) \rangle$ that approximates
$\exp[ ( t_I^{~} - t_F^{~} ) H / i \hbar ] | \Psi( t_F^{~} ) \rangle$. The state
$| \Psi'( t_I^{~} ) \rangle$ thus obtained is not actually the same as the initial
state $| \Psi( t_I^{~} ) \rangle$. This is an example of the
slight asymmetry in time in the real-time DMRG method. Qualitatively 
speaking, the discrepancy between $| \Psi( t_I^{~} ) \rangle$ 
and $| \Psi'_{~}( t_I^{~} ) \rangle$ can be attributed to the accumulation of 
truncation error caused by the 
repeated use of time adopted renormalization processes. 

In order to recover the time-reversal symmetry, we 
introduce a kind of least action principle, which is related to MPSs on all the
time slices. Also we intend to prevent the accumulation of truncation error. 
For these purposes we employ a functional $I = \int {\cal L} dt$, which is not only
stationary but is also minimum for the actual time evolution from 
$| \Psi( t_I^{~} ) \rangle$ to  $| \Psi( t_F^{~} ) \rangle $. 
Minimization for the integral of the
Lagrangian like function ${\cal L}$, which is bilinear in $| \Psi( t ) \rangle$, 
 for the time span $t_I^{~} < t < t_F^{~}$ draws a way of improving  MPS on each 
time slice iteratively. In a sense the optimization process that we explain in the
following can be regarded as `the
finite-time DMRG algorithm', which sweeps MPS between $t_I^{~}$ to $t_F^{~}$ iteratively.
In contrast to the spacial sweeping process in the finite-size DMRG algorithm applied
to ground state problems, the sweeping toward and backward the time direction 
can be performed simultaneously by parallel computation.

In the next section, we introduce a kind of action $I = \int {\cal L} dt$ that is simply written 
as the square error with respect to the small time evolution by transfer matrices. 
In \S 3 we explain how the least action principle draws the optimization conditions for MPS
on each time slices. Conclusions are summarized in the last section.

\section{Least Action Principle}

Consider the time evolution of an isolated quantum state in the Sh\"odinger picture
\begin{equation}
 \, \frac{\partial}{\partial t} \, | \Psi( t ) \rangle = \frac{H( t )}{i \hbar} \,  | \Psi( t ) \rangle \, ,
\end{equation}
where $H( t )$ is the Hamiltonian of the system. We have divided the standard formulation 
by $i \hbar$ in order to concentrate
on the time evolution of the quantum state. The formal solution of this equation 
from the initial state $| \Psi( t_I^{~} ) \rangle$ is given by
\begin{equation}
| \Psi( t_F^{~} ) \rangle = \exp \left( \frac{1}{i \hbar} \int_{t_I^{~}}^{t_F^{~}} H( t ) dt \right) | \Psi( t_I^{~} ) \rangle \, ,
\end{equation}
where we assume the time order in the exponential of the integral. Normally the function
\begin{equation}
{\cal L}( t ) = i \hbar \, \langle \Psi( t ) | 
\left( \frac{\partial}{\partial t} - \frac{H( t )}{i \hbar} \right) 
| \Psi( t ) \rangle 
\end{equation}
is chosen as the Lagrangian. If we do not care about the physical dimension of
the Lagrangian, there are actually a lot of functions that draws Eq.(2.1) by way of 
the stationary condition. 
Among such functions, there is non-negative one
\begin{eqnarray}
{\cal L}'( t ) &=& \langle \Psi( t ) | 
\left( \frac{\partial}{\partial t} - \frac{H( t )}{i \hbar} \right)^{\dagger}_{~}
\left( \frac{\partial}{\partial t} - \frac{H( t )}{i \hbar} \right) 
| \Psi( t ) \rangle \nonumber\\
&=& \left| \left(  \frac{\partial}{\partial t} - \frac{H( t )}{i \hbar} \right) 
| \Psi( t ) \rangle \right|^2_{~} \, ,
\end{eqnarray}
which we treat in the following. Though there is no profit of considering ${\cal L}'( t )$ 
instead of ${\cal L}$ in analytical calculations, it is of use for 
finding a variational principle in the real-time DMRG method.
We define a functional $I$, which corresponds to a kind of action from the initial 
time $t_I^{~}$ to the final one $t_F^{~}$, by the integral~\cite{dimension}
\begin{equation}
I = \int_{t_I^{~}}^{t_F^{~}} \left| \left( \frac{\partial}{\partial t} - \frac{H( t )}{i \hbar} \right) 
| \Psi( t ) \rangle \right|^2_{~} dt \, .
\end{equation}
In order to simplify the formulation, we concentrate on time-independent Hamiltonian 
in the following. (Introduction of time dependence is straight forward.)

For the numerical treatment of time evolutions, let us divide the time span into $N$ 
segments and introduce discrete time
$t_\ell^{~} = t_I^{~} + \ell \Delta t$, where $\Delta t = ( t_F^{~} - t_I^{~} ) / N$.
The final state is then formally expressed as
\begin{equation}
| \Psi( t_F^{~} ) \rangle = 
\exp \biggl[ \frac{t_F^{~} - t_I^{~}}{i \hbar} \,  H \biggr] | \Psi( t_I^{~} ) \rangle =
{\cal T}^N_{~} | \Psi( t_I^{~} ) \rangle
\end{equation}
using the short-time evolution operator 
\begin{equation}
{\cal T} = \exp\bigl( \frac{\Delta t}{i \hbar} \, H \bigr) \, .
\end{equation}
In the same manner, the state at $t_{i+1}^{~}$ is written as
\begin{equation}
| \Psi( t_{i+1}^{~} ) \rangle = {\cal T} | \Psi( t_{i}^{~} ) \rangle 
= {\cal T}^{i+1}_{~} | \Psi( t_I^{~} ) \rangle\, .
\end{equation}
As discrete analogues of the functional in Eq.(2.5), we employ an error function
\begin{equation}
I_{f}^{~} = \sum_{i=0}^{N-1} \, \biggl| | \Psi( t_{i+1}^{~} ) \rangle - {\cal T}
| \Psi( t_{i}^{~} ) \rangle \biggl|^2_{~} 
\end{equation}
or the similar one
\begin{equation}
I_{b}^{~} = \sum_{i=0}^{N-1} \, \biggl| {\cal T}^{-1}_{~} | \Psi( t_{i+1}^{~} ) \rangle -  
| \Psi( t_{i}^{~} ) \rangle \biggl|^2_{~} \, ,
\end{equation}
where we have introduced a backward small-time evolution operator
\begin{equation}
{\cal T}^{-1}_{~} = \exp\bigl( - \frac{\Delta t}{i \hbar} \, H \bigr) \, .
\end{equation}
These two error functions are actually the same, since we have
${\cal T}{\cal T}^{-1}_{~} = id$, and thus
\begin{equation}
| \Psi( t_{i+1}^{~} ) \rangle - {\cal T} | \Psi( t_i^{~} ) \rangle = 
{\cal T} \bigl( {\cal T}^{-1}_{~} | \Psi( t_{i+1}^{~} ) \rangle - | \Psi( t_i^{~} ) \rangle \bigr)
\end{equation}
is satisfied.  Let us focus on the minimization of a part of the error
\begin{equation}
\biggl| {\cal T}^{-1}_{~} | \Psi( t_{i+1}^{~} ) \rangle -  
| \Psi( t_{i}^{~} ) \rangle \biggl|^2_{~} + \,\,
\biggl| | \Psi( t_{i}^{~} ) \rangle - {\cal T} 
| \Psi( t_{i-1}^{~} ) \rangle \biggl|^2_{~}
\end{equation}
that is related to $| \Psi( t_i^{~} ) \rangle$. The stationary condition with respect to the variation
$| \Psi_i^{~} \rangle \rightarrow | \Psi_i^{~} \rangle + | \delta \Psi \rangle$ 
draws the optimal condition 
\begin{equation}
| \Psi( t_i^{~} ) \rangle = \frac{1}{2} \biggl[ {\cal T}^{-1}_{~} | \Psi( t_{i+1}^{~} ) \rangle +
{\cal T} | \Psi( t_{i-1}^{~} ) \rangle  \biggr]
\end{equation}
for those states $| \Psi( t_i^{~} ) \rangle$ at $t_i^{~} <  t_F^{~}$, and 
\begin{equation}
| \Psi( t_N^{~} ) \rangle = {\cal T} | \Psi( t_{N-1}^{~} ) \rangle
\end{equation}
for $| \Psi( t_F^{~} ) \rangle$.

As an example of one-dimensional (1D) systems, let us consider a lattice model of 
length $L$, whose Hamiltonian is written as the sum of nearest neighbor interactions
\begin{eqnarray}
H 
&=& \sum_{i=\ell}^{L-1} \, h_{\ell, \ell+1}^{~}
= \sum_{\ell ={\rm odd}}^{~} \, h_{\ell, \ell+1}^{~} 
+ \sum_{\ell ={\rm even}}^{~} \, h_{\ell, \ell+1}^{~}\nonumber\\
&=& H_1^{~} + H_2^{~} \, ,
\end{eqnarray}
where we assume the open boundary condition throughout this article.
In such a case the time evolution is well approximated by the Trotter formula
%~\cite{Trotter}
\begin{eqnarray}
| \Psi( t_F^{~} ) \rangle &=& \biggl[ \exp\bigl( \frac{\Delta t}{i \hbar} \, H_2^{~} \bigr) 
\exp\bigl( \frac{\Delta t}{i \hbar} \, H_1^{~} \bigr) \biggr]^N_{~} | \Psi( t_I^{~} ) \rangle 
\nonumber\\
&=& \biggl[ {\cal T}_2^{~} \, {\cal T}_1^{~} \biggr]^N_{~} | \Psi( t_I^{~} ) \rangle \, ,
\end{eqnarray}
where ${\cal T}_1^{~}$ and ${\cal T}_2^{~}$ are written as product of non-overlapping 
local time evolutions
\begin{eqnarray}
{\cal T}_1^{~} 
&=& \prod_{\ell={\rm odd}}^{~} \exp\bigl( \frac{\Delta t}{i \hbar} \, h_{\ell,\ell+1}^{~} \bigr) 
= \prod_{\ell={\rm odd}}^{~} \tau_{\ell,\ell+1}^{~} \, ,
\nonumber\\
{\cal T}_2^{~} 
&=& \prod_{\ell={\rm even}}^{~} \exp\bigl( \frac{\Delta t}{i \hbar} \, h_{\ell,\ell+1}^{~} \bigr) 
= \prod_{\ell={\rm even}}^{~} \tau_{\ell,\ell+1}^{~} \, .
\end{eqnarray}
The Trotter decomposition introduces a new state 
$| \phi_{i+1/2}^{~} \rangle \equiv {\cal T}_1^{~} | \Psi( t_i^{~} ) \rangle$ between 
$| \Psi_i^{~} \rangle \equiv | \Psi( t_i^{~} ) \rangle$ and 
$| \Psi_{i+1}^{~} \rangle \equiv | \Psi( t_{i+1}^{~} ) \rangle$.
The index $i + 1/2$ of $| \phi_{i+1/2}^{~} \rangle$ just means that it 
is between $i$ and $i + 1$; 
note that $| \phi_{i+1/2}^{~} \rangle$ 
does not correspond to $| \Psi( t_{i}^{~} + \Delta t / 2 ) \rangle$. 
Now we have $2N$ numbers of states, and the error function in Eqs.(2.9) or (2.10) is modified as
\begin{eqnarray}
I &=& \sum_{i=0}^{N-1}  \left| | \phi_{i+1/2}^{~} \rangle - {\cal T}_1^{~} | \Psi_i^{~} \rangle 
\right|^2_{~} \nonumber\\
&+& \sum_{i=0}^{N-1}  \left| | \Psi_{i+1}^{~} \rangle - {\cal T}_2^{~} | \phi_{i+1/2}^{~} \rangle
\right|^2_{~} \, .
\end{eqnarray}
As we have done for Eqs.(2.13) and (2.14), 
minimization of this error function draws the following conditions
\begin{eqnarray}
| \Psi_i^{~} \rangle &=& \frac{1}{2} \left[ 
{\cal T}_2^{~} | \phi_{i-1/2}^{~} \rangle +
\left( {\cal T}_1^{~} \right)^{-1}_{~} | \phi_{i+1/2}^{~} \rangle
\right] \, ,\nonumber\\
| \phi_{i+1/2}^{~} \rangle &=& \frac{1}{2} \left[ 
{\cal T}_1^{~} | \Psi_i^{~} \rangle +
\left( {\cal T}_2^{~} \right)^{-1}_{~} | \Psi_{i+1}^{~} \rangle
\right]
\end{eqnarray}
for $i < N$. Only at the final time
\begin{equation}
| \Psi_N^{~} \rangle =
{\cal T}_2^{~} | \phi_{N-1/2}^{~} \rangle
\end{equation}
should be satisfied. From equation (2.20), it is apparent that 
we can deal $| \phi_{i+1/2}^{~} \rangle$ equivalently with $| \Psi_i^{~} \rangle$ as 
long as the minimization of the error function $I$ is concerned. We
therefore explain the optimization of $| \Psi_i^{~} \rangle$ only in the following.
This variational formulation is time-symmetric, in the sense that 
both ${\cal T}_1^{~}$ and  ${\cal T}_2^{~}$ are 
invertible and thus $I_b^{~} = I_f^{~}$.

\section{Optimization of Matrix Product States}

In the framework of DMRG method, all the states are (implicitly) written as the MPS.
Let us write $| \Psi^{~}_{i} \rangle = | \Psi( t^{~}_{i} ) \rangle$ in the form of the orthogonal 
MPS~\cite{Ostlund,Takasaki,Shoe}
\begin{equation}
B[ s_1^{~} ] \ldots B[ s_\ell^{~} ] \,\, \Lambda^{\rm B}_{~}
\,\, B[ s_{\ell+1}^{~} ] \ldots B[ s_L^{~} ] \, | s_1^{~} \ldots s_L^{~} \rangle \, ,
\end{equation}
which corresponds to the division of the lattice into the $\ell$-site left part and the
$L - \ell$ site right part. The notation $s_j^{~}$ represent the site variable --- say, the spin
variable --- at position $j$. Suppose that the degree of freedom of each $s_j^{~}$ is $d$. 
Since we are treating a system with open boundary condition, 
$B[ s_1^{~} ]$ is a $1 \times d$ matrix (= row vector) 
\begin{equation}
B_{\times \alpha}^{~}[ s_1^{~} ] = \delta( \alpha, s_1^{~} )
\end{equation}
and $B[ s_L^{~} ]$ is a $d \times 1$ matrix (= column vector)
\begin{equation}
B_{\alpha \times}^{~}[ s_L^{~} ] = \delta( \alpha, s_L^{~} ) \, ,
\end{equation}
where the matrix indices $`\times'$ and $\alpha$ represent the 1-state 
and $d$-state auxiliary variables, respectively.
In the following we consider the case where the size of the matrix
is less than a fixed integer $m$, which is the maximum number of states kept in the DMRG
method.
Figure 1 shows the structure of the orthogonal MPS. The white circles represent 
spin variables
$s_i^{~}$, and the black squares represent auxiliary variables whose sum is taken over.
The triangles correspond to the orthogonal matrices $B[ s_j^{~} ]$. The large circle at the 
dividing point of the system represent the diagonal matrix $\Lambda^{\rm B}_{~}$.

\begin{figure}
\epsfxsize=75mm 
\centerline{\epsffile{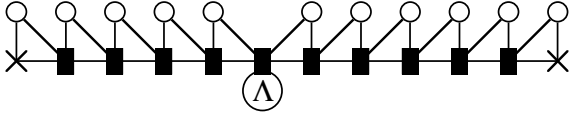}}
\caption{Graphical representation of the MPS in Eq.~(3.1) for the case $L = 12$.}
\label{fig:1}
\end{figure}

The matrices $B[ s_j^{~} ]$ in the left side of $\Lambda^{\rm B}_{~}$ satisfies 
the orthogonal relation
\begin{equation}
\sum_{s_j^{~} \alpha}^{~} 
\left( B^{~}_{\alpha \beta}[ s_j^{~} ] \right)^*_{~}
B^{~}_{ \alpha \gamma}[ s_j^{~} ] = \delta_{\beta \gamma}^{~} ~~~~~( j \le \ell ) \, ,
\end{equation}
and that in the right side  of $\Lambda^{\rm B}_{~}$ satisfies 
\begin{equation}
\sum_{s_j^{~} \beta}^{~} 
\left( B^{~}_{\alpha \beta}[ s_j^{~} ] \right)^*_{~}
B^{~}_{ \gamma \beta}[ s_j^{~} ] = \delta_{\alpha \gamma}^{~} ~~~~~( j > \ell ) \, .
\end{equation}
The diagonal matrix $\Lambda^{\rm B}_{~}$ contains the singular values 
 $\lambda^{\rm B}_{\alpha}$ with respect to the left-right division of the 
state $| \Psi_i^{~} \rangle$ at the position $\ell$. The matrix $\Lambda^{\rm B}_{~}$ is
dependent to $\ell$, and satisfies the condition
\begin{equation}
{\rm Tr} \left( \Lambda^{\rm B}_{~} \right)^2_{~} = 
\sum_{\alpha}^{~} \left( \lambda^{\rm B}_{\alpha} \right)_{~}^2 = 1 
\end{equation}
when $| \Psi_i^{~} \rangle$ is normalized.
The position of the left-right division can be shifted arbitrarily by way of
the relation~\cite{Ostlund,PWFRG,Shift,Takasaki}
\begin{equation}
B[ s_\ell^{~} ] \, \Lambda^{\rm B}_{~} = 
{\tilde \Psi}_i^{~}[ s_\ell^{~} ] = 
\Lambda^{\rm B}_{~} \, B[ s_\ell^{~} ] \, ,
\end{equation}
where $B[ s_\ell^{~} ]$ of  $B[ s_\ell^{~} ] \, \Lambda^{\rm B}_{~}$ satisfies Eq.(3.4), 
and that of  $\Lambda^{\rm B}_{~} \, B[ s_\ell^{~} ]$ satisfies Eq.(3.5). 
(See Fig.~2.) Using 
the {\it renormalized wave function} ${\tilde \Psi}_i^{~}[ s_\ell^{~} ]$ defined in the
above equation, we can express $| \Psi_i^{~} \rangle$ as
\begin{equation}
B[ s_1^{~} ] \ldots B[ s_{\ell-1}^{~} ] \,\, {\tilde \Psi}_i^{~}[ s_\ell^{~} ] 
\,\, B[ s_{\ell+1}^{~} ] \ldots B[ s_L^{~} ] \, | s_1^{~} \ldots s_L^{~} \rangle \, .
\end{equation}

\begin{figure}
\epsfxsize=50mm 
\centerline{\epsffile{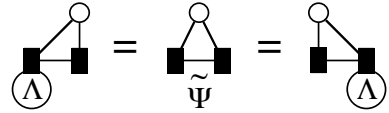}}
\caption{Renormalized wave function defined in Eq.(3.7).}
\label{fig:2}
\end{figure}

Because of the orthogonality in Eqs.(3.4) and (3.5), the inner product 
$\langle \Psi_i^{~} | \Psi_i^{~} \rangle$ can be expressed simply as 
\begin{eqnarray}
\langle \Psi_i^{~} | \Psi_i^{~} \rangle 
&=&
\sum_{\alpha \beta s_\ell^{~}} 
\left( {\tilde \Psi}_{\alpha \beta}^{~}[ s_\ell^{~} ] \right)^*_{~}
{\tilde \Psi}_{\alpha \beta}^{~}[ s_\ell^{~} ] \nonumber\\
&=&
\sum_{\alpha \beta s_\ell^{~}} 
\left( B^{~}_{\alpha \beta}[ s_\ell^{~} ] \, \lambda^{\rm B}_{\beta} \right)^*_{~}
\left( B^{~}_{\alpha \beta}[ s_\ell^{~} ] \, \lambda^{\rm B}_{\beta} \right) 
\nonumber\\
&=& \sum_{\beta}^{~} \left( \lambda^{\rm B}_{\beta} \right)^2_{~} = 1 \, .
\end{eqnarray}
The right hand side of the first line can be regarded as the norm of 
the renormalized wave function ${\tilde \Psi}_i^{~}[ s_\ell^{~} ]$, which we express
as $( {\tilde \Psi}_i^{~} | {\tilde \Psi}_i^{~} )$ in the following.

In the same manner as $| \Psi_i^{~} \rangle$, the states
$| \phi^{~}_{i-1/2} \rangle$ and $| \phi^{~}_{i+1/2} \rangle$ can be
expressed in the form of the orthogonal MPS
\begin{eqnarray}
| \phi^{~}_{i-1/2} \rangle &=& 
A[ s_1^{~} ]  \ldots A[ s_\ell^{~} ] \,\, \Lambda^{\rm A}_{~}
\,\, A[ s_{\ell+1}^{~} ] \ldots A[ s_L^{~} ] \, | s_1^{~} \ldots s_L^{~} \rangle
\nonumber\\
| \phi^{~}_{i+1/2} \rangle &=& 
C[ s_1^{~} ]  \ldots C[ s_\ell^{~} ] \,\, \Lambda^{\rm C}_{~}
\,\, C[ s_{\ell+1}^{~} ] \ldots C[ s_L^{~} ] \, | s_1^{~} \ldots s_L^{~} \rangle \, ,
\nonumber\\
\end{eqnarray}
respectively, 
and the corresponding renormalized wave functions are
${\tilde \phi}^{~}_{i-1/2}[ s_\ell^{~} ]  = A[ s_\ell^{~} ] \, \Lambda^{\rm A}_{~}$ and
${\tilde \phi}^{~}_{i+1/2}[ s_\ell^{~} ]  = C[ s_\ell^{~} ] \, \Lambda^{\rm C}_{~}$.

Consider a variation of $| \Psi_i^{~} \rangle$ with respect to the local
change in MPS caused by ${\tilde \Psi}[ s_\ell^{~} ] \rightarrow 
{\tilde \Psi}[ s_\ell^{~} ] + \delta {\tilde \Psi}[ s_\ell^{~} ]$. 
What should be minimized with respect to this variation is
\begin{eqnarray}
&&
\left| \bigl( {\cal T}_1^{~} \bigr)^{-1}_{~} 
| \phi_{i+1/2}^{~} \rangle -  | \Psi_i^{~} \rangle \right|^2_{~} + 
\left| {\cal T}_2^{~} | \phi_{i-1/2}^{~} \rangle -  | \Psi_i^{~} \rangle \right|^2_{~}
\nonumber\\
&=& 
\langle \Psi_i^{~} | \Psi_i^{~} \rangle - 
\langle \phi_{i+1/2}^{~} | {\cal T}_1^{~} | \Psi_i^{~} \rangle -
\langle \phi_{i-1/2}^{~} | \bigl( {\cal T}_2^{~} \bigr)^{-1}_{~}  | \Psi_i^{~} \rangle +
1 \nonumber\\
&& + \,\,\, {\rm h.c.} ~~ .
\end{eqnarray}
Using the matrix product structures of each state, the inner products
$\langle \phi_{i + 1/2}^{~} | {\cal T}_1^{~} | \Psi_i^{~} \rangle$ 
 and $\langle \phi_{i - 1/2}^{~} | \left( {\cal T}_2^{~} \right)^{-1}_{~} | \Psi_i^{~} \rangle$
can be calculated
rapidly of the order of $m^3_{~} L$ in computational time.~\cite{Verst_1,Verst}
To simplify the notation let us introduce new renormalized wave functions
${\tilde \Psi}^{-}_{~}[ s_\ell^{~} ]$ and 
${\tilde \Psi}^{+}_{~}[ s_\ell^{~} ]$
that satisfies the relations
\begin{eqnarray}
( {\tilde \Psi}^{-}_{~} | {\tilde \Psi}_i^{~} ) 
&=&
\sum_{\alpha \beta s_\ell^{~}}^{~} \left( {\tilde \Psi}^{-}_{\alpha \beta}[ s_\ell^{} ]
\right)^*_{~} {\tilde \Psi}_{\alpha \beta}^{~}[ s_\ell^{~} ] \nonumber\\
&=&
\langle \phi_{i - 1/2}^{~} | \left( {\cal T}_2^{~} \right)^{-1}_{~} | \Psi_i^{~} \rangle 
\, ,\nonumber\\
( {\tilde \Psi}^{+}_{~} | {\tilde \Psi}_i^{~} ) 
&=&
\sum_{\alpha \beta s_\ell^{~}}^{~} \left( {\tilde \Psi}^{+}_{\alpha \beta}[ s_\ell^{} ]
\right)^*_{~} {\tilde \Psi}_{\alpha \beta}^{~}[ s_\ell^{~} ] \nonumber\\
&=&
\langle \phi_{i + 1/2}^{~} | {\cal T}_1^{~} | \Psi_i^{~} \rangle \, .
\end{eqnarray}
It is easily shown that the computational cost of obtaining both
${\tilde \Psi}^{-}_{~}[ s_\ell^{~} ]$ and 
${\tilde \Psi}^{+}_{~}[ s_\ell^{~} ]$
is also of the order of $m^3_{~} L$.

\begin{figure}
\epsfxsize=60mm
\centerline{\epsffile{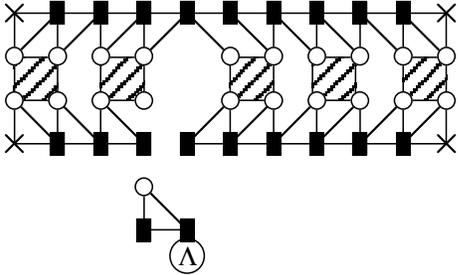}}
\caption{Renormalized wave functions ${\tilde \Psi}^{-}_{~}[ s_\ell^{~} ]$ (upper)
and ${\tilde \Psi}_i^{~}[ s_\ell^{~} ]$ (lower). Squares represent the local evolution
operator $\tau_{j, j+1}^{~}$ contained in ${\cal T}_2^{~}$.}
\label{fig:3}
\end{figure}

The quantity in Eq.~(3.11) can be then written in short
\begin{equation}
\left\{ ( {\tilde \Psi}_i^{~} | {\tilde \Psi}_i^{~} ) 
- ( {\tilde \Psi}^{+}_{~} | {\tilde \Psi}_i^{~} ) 
- ( {\tilde \Psi}^{-}_{~} | {\tilde \Psi}_i^{~} ) + 1 \right\} + \, {\rm h. c.} \, ,
\end{equation}
and we finally obtain the optimal condition
\begin{equation}
{\tilde \Psi}_i^{~}[ s_\ell^{~} ] = \frac{1}{2} 
{\tilde \Psi}_{~}^-[ s_\ell^{~} ] + \frac{1}{2} {\tilde \Psi}_{~}^+[ s_\ell^{~} ] 
\end{equation}
that improves the renormalized wave function ${\tilde \Psi}[ s_\ell^{~} ]$.
Only at the final time $t_F^{~} = t_N^{~}$ the condition is modified as
\begin{equation}
{\tilde \Psi}_N^{~}[ s_\ell^{~} ] = \frac{1}{2} 
{\tilde \Psi}_{~}^-[ s_\ell^{~} ] \, .
\end{equation}

Even when the states $| \Psi_i^{~} \rangle$ and $| \phi_{i\pm 1/2}^{~} \rangle$ are 
normalized, the improved ${\tilde \Psi}_i^{~} [ s_\ell^{~} ]$ created by Eqs.(3.14) and (3.15) 
does not always satisfy the normalization condition $( {\tilde \Psi}_i^{~} | {\tilde \Psi}_i^{~} ) = 1$. 
Thus we normalize ${\tilde \Psi}_i^{~} [ s_\ell^{~} ]$  as
\begin{equation}
{\tilde \Psi}_i^{~}[ s_\ell^{~} ]  / ( {\tilde \Psi}_i^{~} | {\tilde \Psi}_i^{~} )^{1/2}_{~} 
\rightarrow {\tilde \Psi}_i^{~}[ s_\ell^{~} ] \, 
\end{equation}
after each local optimization.
Schmidt orthogonalization of thus improved ${\tilde \Psi}_i^{~}[ s_\ell^{~} ]$ gives
improved orthogonal matrix $B[ s_\ell^{~} ]$ and the singular value $\Lambda^{\rm B}_{~}$ 
at the position $\ell$ on the time slice $t_i^{~}$. 
Sweeping this improving process from $i = 1$ to $i = L$ 
for several times, one obtains optimized $| \Psi_i^{~} \rangle$ with respect to 
the fixed $| \phi_{i\pm1/2}^{~} \rangle$.~\cite{Verst_1,Verst}

To perform such a sweeping process for all the time slices requires a huge amount of
numerical calculation. But this is not totally unrealistic, since improvements of 
$| \Psi_i^{~} \rangle = | \Psi( t_i^{~})  \rangle$ and 
$| \Psi_j^{~} \rangle = | \Psi( t_j^{~} \neq t_i^{~} )  \rangle$ can be performed 
independently under the condition that $| \phi_{i\pm1/2} \rangle$ for every $i$ is fixed. 
Also we can say that $| \phi_{i\pm1/2} \rangle$ can be improved simultaneously
for every $i$ when all the $| \Psi_i^{~} \rangle$ are fixed. The nature enables us to
perform
the parallel computation. After improving $| \Psi_i^{~} \rangle$ and
$| \phi_{i\pm1/2} \rangle$ on all the time slices alternatively for numbers of times, 
we obtain the matrix product states that minimizes the error function in (2.19).~\cite{fix}
A more realistic way of calculation is to boost the state from $| \Psi( t_N^{~} ) \rangle$ to
$| \Psi( t_{N+1}^{~} ) \rangle$ using the conventional real-time DMRG methods,
and to optimize the obtained MPSs for $M (<\!< N)$ numbers of time slices from the frontier
after each time boost.

The explained procedure does not change the size $m$ of 
each matrix. 
Occasionally it is better to change the value of $m$ site by site
according to the truncated weights in the renormalization process.
Variation with respect to the extended 
renormalized wave function
\begin{equation}
{\tilde \Psi}_i^{~}[ s_\ell^{~} s_{\ell+1}^{~} ] = 
B[ s_\ell^{~} ] \, \Lambda^{\rm B}_{~} B[ s_{\ell+1}^{~} ] = 
B[ s_\ell^{~} ]  B[ s_{\ell+1}^{~} ] \, \Lambda^{\rm B}_{~}
\end{equation}
makes it possible to adjust $m$ dynamically during the calculation. 

\begin{figure}
\epsfxsize=70mm 
\centerline{\epsffile{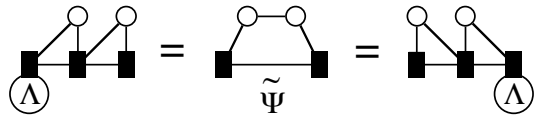}}
\caption{Extended renormalized wave function in Eq.(3.17).}
\label{fig:2}
\end{figure}

\section{Conclusion and discussion}

We reformulate the real-time DMRG method so that it minimizes a kind of
discrete action, which corresponds to the square of error
in the numerical time evolution. As a result the time symmetry is
recovered from the level of variational formulation. 
The minimization process can be 
performed via parallel computation. 

Let us discuss the origin of the slight time asymmetry in the conventional
real-time DMRG algorithms. Consider the multiplication of ${\cal T}_1^{~}$
to $| \Psi_i^{~} \rangle$ that is expressed as the MPS. It is possible to represent 
${\cal T}_1^{~} | \Psi_i^{~} \rangle$ exactly in the form of MPS again, but this 
requires more degrees of freedom than the original MPS representation of
$| \Psi_i^{~} \rangle$. Therefore a truncation ($=$ renormalization) process $R$
is applied to ${\cal T}_1^{~} | \Psi_i^{~} \rangle$ to keep the degree of freedom
nearly constant. As a result, very small but non-zero discrepancy 
\begin{equation}
 {\cal T}_1^{~} | \Psi_i^{~} \rangle - R {\cal T}_1^{~} | \Psi_i^{~} \rangle
\end{equation}
arises. Since the truncation process satisfies $R^2_{~} = R$, and since
$| \Psi_i^{~} \rangle$ is obtained by applying $R$ to 
${\cal T}_2^{~} | \phi_{i-1/2}^{~} \rangle$, we obtain $R \, | \Psi_i^{~} \rangle = 
| \Psi_i^{~} \rangle$. Thus the above discrepancy can by written as
\begin{equation}
{\cal T}_1^{~} R \, | \Psi_i^{~} \rangle - R {\cal T}_1^{~} | \Psi_i^{~} \rangle =
[ {\cal T}_1^{~}, \, R ] \, | \Psi_i^{~} \rangle \, .
\end{equation}
This tiny error cannot be recovered by the 
multiplication of $\bigl( {\cal T}_1^{~} \bigr)^{-1}_{~}$ from the left,
since $\bigl( {\cal T}_1^{~} \bigr)^{-1}_{~} R \, {\cal T}_1^{~}$ is not the 
identity operator. For ${\cal T}_2^{~}$ there is the same kind of 
discrepancy $[ {\cal T}_2^{~}, \, R ] \, | \phi_{i-1/2}^{~} \rangle$. 
Although these are tiny errors, repeated use of the truncation $R$ after 
each time evolution may introduce exponentially growing numerical error.
The variational treatment presented in this article recovers the
time symmetry by considering Eq.(3.11). The time boost at $t_F^{~}$ is
still asymmetric as shown in Eq.(3.15), and this asymmetry should be
removed afterward through the repeated optimization processes.

In our formulation, we assumed that the Hamiltonian can be decomposed
into two non-overlapping parts. For the cases where there are long-range
interactions, the multi-targeting scheme by Feiguin and White is 
of use.~\cite{D_White} The outline is to optimize $| \Psi( t_{i}^{~} ) \rangle$ 
so that the error function in Eq.(2.13) is minimized. This process is realized by
rewriting the involved states $| \Psi( t_{i-1}^{~} ) \rangle$, 
$| \Psi( t_{i}^{~} ) \rangle$, and $| \Psi( t_{i+1}^{~} ) \rangle$ using the
same orthogonal matrices.~\cite{Jeck} It is interesting that the method of Feiguin and White 
performs the same kind of optimization for the succeeding 4 states 
$| \Psi( t_{i}^{~} ) \rangle$, 
$| \Psi( t_{i+1/3}^{~} ) \rangle$, $| \Psi( t_{i+2/3}^{~} ) \rangle$, and 
$| \Psi( t_{i+1}^{~} ) \rangle$.~\cite{asym2}

Remember that there are many possible choices of Lagrangian-like function 
even in continuous time formulations, and there are much more for the discrete time cases.
The authors have just considered one of them.
Looking at the classical simulation of Newton equation, there are
various techniques that would be implemented in real-time DMRG algorithm. For example,
automatic adjustment of $\Delta t$, use of the symplectic structure, 
position dependent choice of $\Delta t$, etc. It is worth considering
what kind of Lagrangian or error function exist toward these extensions.

The variational method introduced in this article can be applied for
transfer matrix problems in two-dimensional classical lattice 
models, if $T^{-1}_{~}$ is represented as a product of local factors. 
In the same manner, classical simulations of quantum circuits~\cite{Vidal2,Kawaguchi}
are in our scope, since the transfer matrices that represent quantum
operations are always invertible. For those cases where there is no
inverse, or non-local terms appears in $T^{-1}_{~}$, we still 
do not obtain an appropriate variational functional.

T. N. is partially supported by a Grant-in-Aid for Scientific Research from the Ministry of 
Education, Science, Sports and Culture.


\begin{thebibliography}{99}
\bibitem{dimension}Note that the integral does not have the proper dimension of the
action. It is just the functional integral of some function ${\cal L}'( t )$.
\bibitem{fix}After the complete improvement, ${\tilde \Psi}^{\pm}_{~}[ s_\ell^{~} ]$ is very 
close to ${\tilde \Psi}[ s_\ell^{~} ]$. They become the same both in the limit 
$\Delta t \rightarrow 0$ or for sufficiently large $m$.
\bibitem{asym2}After expressing the succeeding 3 states by use of the same orthogonal
matrices, there is no time asymmetry in the time boost in Eq.~(2.8). However, 
very few but nonzero error enters during the rewriting process of MPS, which 
plays the same role as $R$ in Eqs.(4.1) and (4.2).
\end{thebibliography}
\end{document}